\documentclass[twocolumn,showpacs,preprintnumbers,amsmath,amssymb]{revtex4}

\usepackage{graphicx}
\usepackage{dcolumn}
\usepackage{bm}

\newcommand{\eigreim}[6]  
{\begin{picture}(#1,#2)(#3,#4)
\put(0,0){\line(1,0){#5}}
\put(0,0){\line(0,1){#5}}
\put(0,0){\line(-1,0){#5}}
\put(0,0){\line(0,-1){#5}}
\put(#6,0.0){\circle*{#6}}
\put(-#6,0.0){\circle*{#6}}
\put(0.0,#6){\circle*{#6}}
\put(0.0,-#6){\circle*{#6}}
\end{picture}
}

\newcommand{\eigreimout}[6]  
{\begin{picture}(#1,#2)(#3,#4)
\put(0,0){\line(1,0){#5}}
\put(0,0){\line(0,1){#5}}
\put(0,0){\line(-1,0){#5}}
\put(0,0){\line(0,-1){#5}}
\put(-#6,#6){\circle*{#6}}
\put(#6,#6){\circle*{#6}}
\put(-#6,-#6){\circle*{#6}}
\put(#6,-#6){\circle*{#6}}
\end{picture}
}

\newcommand{\eigrenegnegpospos}[6]  
{\begin{picture}(#1,#2)(#3,#4)
\put(0,0){\line(1,0){#5}}
\put(0,0){\line(0,1){#5}}
\put(0,0){\line(-1,0){#5}}
\put(0,0){\line(0,-1){#5}}
\put(-1.2,0.0){\circle*{#6}}
\put(-#5,0.0){\circle*{#6}}
\put(#5,0.0){\circle*{#6}}
\put(1.2,0.0){\circle*{#6}}
\end{picture}
}

\newcommand{\eigimnegnegpospos}[6]  
{\begin{picture}(#1,#2)(#3,#4)
\put(0,0){\line(1,0){#5}}
\put(0,0){\line(0,1){#5}}
\put(0,0){\line(-1,0){#5}}
\put(0,0){\line(0,-1){#5}}
\put(0.0,-1.2){\circle*{#6}}
\put(0.0,-#5){\circle*{#6}}
\put(0.0,#5){\circle*{#6}}
\put(0.0,1.2){\circle*{#6}}
\end{picture}
}

\newcommand{\eigrenegpos}[6]  
{\begin{picture}(#1,#2)(#3,#4)
\put(0,0){\line(1,0){#5}}
\put(0,0){\line(0,1){#5}}
\put(0,0){\line(-1,0){#5}}
\put(0,0){\line(0,-1){#5}}
\put(-3,-1){{\large $\times$}}
\put(0,-1){{\large$\times$}}
\end{picture}
}

\newcommand{\eigimnegpos}[6]  
{\begin{picture}(#1,#2)(#3,#4)
\put(0,0){\line(1,0){#5}}
\put(0,0){\line(0,1){#5}}
\put(0,0){\line(-1,0){#5}}
\put(0,0){\line(0,-1){#5}}
\put(-1.65,0.5){{\large $\times$}}
\put(-1.65,-2.5){{\large $\times$}}
\end{picture}
}

\newcommand{\eigzerorenegpos}[7]  
{\begin{picture}(#1,#2)(#3,#4)
\put(0,0){\line(1,0){#5}}
\put(0,0){\line(0,1){#5}}
\put(0,0){\line(-1,0){#5}}
\put(0,0){\line(0,-1){#5}}
\put(-1.65,-1){{\large $\times$}}
\put(-2,0.0){\circle*{#6}}
\put(2,0.0){\circle*{#6}}
\end{picture}
}

\newcommand{\eigzeroimnegpos}[7]  
{\begin{picture}(#1,#2)(#3,#4)
\put(0,0){\line(1,0){#5}}
\put(0,0){\line(0,1){#5}}
\put(0,0){\line(-1,0){#5}}
\put(0,0){\line(0,-1){#5}}
\put(-1.65,-1){{\large $\times$}}
\put(0.0,-2){\circle*{#6}}
\put(0.0,2){\circle*{#6}}
\end{picture}
}

\newcommand{\ofour}[7]  
{\begin{picture}(#1,#2)(#3,#4)
\put(0,0){\line(1,0){#5}}
\put(0,0){\line(0,1){#5}}
\put(0,0){\line(-1,0){#5}}
\put(0,0){\line(0,-1){#5}}
\put(-2.3,-1.5){{\LARGE $\times$}}
\put(0.0,0.0){\circle{#6}}
\end{picture}
}

\begin{document}


\title{Stability Analysis of the Lugiato-Lefever Model for Kerr Optical Frequency Combs. \\
       Part I: Case of Normal Dispersion}

\author{Cyril Godey$^1$, Irina Balakireva$^2$,  Aur\'elien Coillet$^2$ and Yanne K. Chembo${^2}$}
\thanks{Corresponding author. E-mail: yanne.chembo@femto-st.fr}
\affiliation{$^1$University of Franche-Comt\'e,  Department of Mathematics [CNRS UMR6623], \\
             16 Route de Gray, 25030 Besan\c con cedex, France. \\
             $^2$FEMTO-ST Institute [CNRS UMR6174], Optics Department, \\
             16 Route de Gray, 25030 Besan\c con cedex, France.}
\date{\today}

\begin{abstract}
We propose a detailed stability analysis of the Lugiato-Lefever model for Kerr optical frequency combs in whispering gallery mode resonators pumped in the normal dispersion regime.
We analyze the spatial bifurcation structure of the stationary states depending on two parameters that are experimentally tunable, namely the pump power and the cavity detuning.
Our study demonstrates that the non-trivial equilibria play an important role in this bifurcation map, as their associated eigenvalues undergo critical bifurcations that are foreshadowing the existence of localized spatial structures.
In particular, we show that in the normal dispersion regime, dark cavity solitons can emerge in the system, and thereby generate a Kerr comb. We also show how these solitons can coexist in the resonator as long as they do not interact with each other. The Kerr combs created by these (sets of) dark solitons are also analyzed, and their stability is discussed as well.
\end{abstract}

\maketitle

\section{Introduction}
\label{intro}

A Kerr combs is a set of equidistant spectral components generated
through the optical pumping of an ultra-high quality factor whispering gallery
mode (WGM) resonator with Kerr nonlinearity. In this system, the WGM resonator is pumped by a continuous-wave (CW) laser, and the pump photons are transferred through four-wave mixing (FWM) to neighboring  cavity eigenmodes. All these excited modes are coupled through FWM, and as a result, may excite an even greater number of modes. This process can generate as much as several hundred oscillating modes, which will be the spectral components constituting the so-called Kerr comb. 

Kerr comb generators are characterized by their conceptual simplicity, structural robustness,
small size, and low power consumption. They are therefore promising candidates to replace femtosecond mode-locked lasers in applications where these features are of particular relevance.  
The theoretical understanding of Kerr comb generation in whispering gallery
mode resonators is currently the focus of a worldwide activity that is
motivated by the wide range of potential applications~\cite{Review_Kerr_combs_Science,Nature_Ferdous,Nature_Selectable_freq,PRA_Scott2,PRL_NIST,Nature_Universal,Nature_Mid_IR,Jove}.
Another incentive is the
necessity to understand the complex light-matter interactions that are induced by the strong confinement of long-lifetime photons in nonlinear media.

Early models for Kerr comb generation were based on a modal expansion approach~\cite{Maleki_PRL_LowThres,YanneNanPRL,YanneNanPRA}, which uses a large set of coupled nonlinear
ordinary differential equations to track the individual dynamics of the WGMs.  
This formalism enabled to determine threshold phenomena, to explain the role of dispersion, as well as some of the mechanisms leading to Kerr comb generation. However, this modal model becomes difficult to analyze theoretically when the number of excited modes is higher than five. 
Indeed, the modal approach simultaneously looses computational cost-effectiveness because the duration of the numerical simulations becomes prohibitively large, as it increases sharply with the number of modes involved in the FWM process~\cite{YanneNanPRA}.

An alternative paradigm has been introduced recently, and it is based on the fact that in the system under study, light circumferentially propagates inside the resonator and can be treated as if it was propagating along an unfolded trajectory with periodic boundary conditions~\cite{GaetaOE,Matsko_OL_2,PRA_Yanne-Curtis,Coen}. In this case, the system can be modelized by a spatiotemporal formalism known as the Lugiato-Lefever equation (LLE), which is a nonlinear Schr\"odinger equation (NLSE) with damping, detuning and driving~\cite{LL}. 
The variable of the LLE is the overall intracavity field, which is the sum of the modal fields described by the modal model.  Equivalence between the modal approach and the spatiotemporal formalism have been demonstrated recently, and enables to understand the Kerr comb generation process from different viewpoints: the modal approach is useful to investigate threshold phenomena when only few modes are involved, while the spatiotemporal formalism is suitable when a very large number of interacting modes are involved~\cite{YanneNanPRA}.
In this latter case, mode-locking between the modes can lead to the formation of narrow pulses such as cavity solitons for example.
From a more general perspective, the LLE formalism shows that Kerr combs are the spectral signature of dissipative patterns or localized structures in WGM resonators~\cite{IEEE_PJ}. 

\begin{figure}
\begin{center}
\includegraphics[width=8cm]{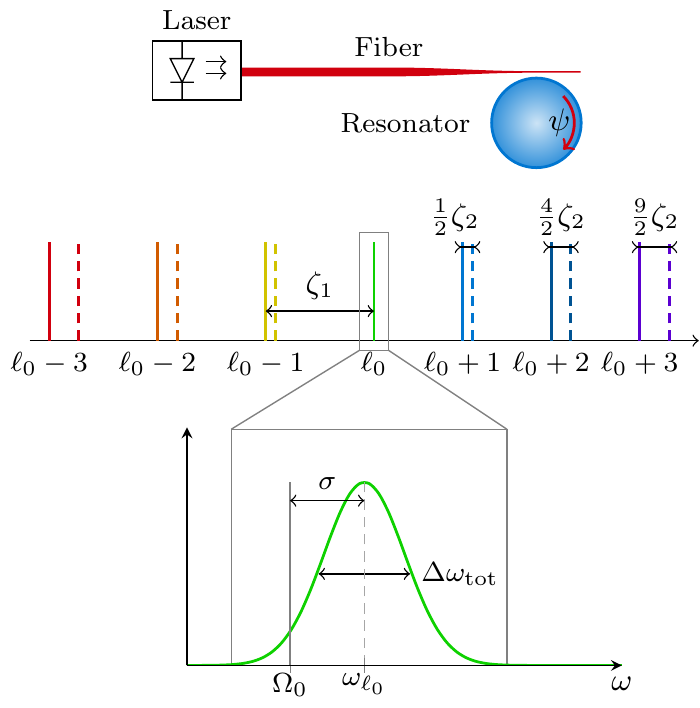}
\end{center}
\caption[Figure1]
{\label{Figure1} (Color online) 
(a) Schematical representation of a WGM resonator.
The pump radiation originates from a CW laser and the coupling is achieved (for example) using a tapered fiber.
(b) Eigenmodes of a WGM resonator. The real location of the modes with normal dispersion is represented in solid lines, while the dashed lines represent the location of the modes if the dispersion was null (perfect equidistance). The zoom-in of the pumped mode $\ell_0$ shows the relationship between linewidth $\Delta \omega_{\rm tot}$, resonance frequency $\omega_{\ell_0}$, laser frequency $\Omega_0$ and cavity detuning $\sigma = \Omega_0 -\omega_{\ell_0}$. Note that the normal dispersion pulls the modes leftwards (redshift).
}
\end{figure}

Amongst the various parameters that are relevant to understand Kerr comb generation, group velocity  dispersion (GVD) is one of the most important. GVD in WGM resonators can be either normal or anomalous, and the real-valued parameter corresponding to GVD have opposite signs depending on the dispersion regime.
From a mathematical point of view, the solutions to be expected using the constitutive dynamical equations are therefore different; from the physics standpoint, the phenomenology is intrinsically different as well. However, the role of dispersion in Kerr comb formation is still a wide open problem, and the various solutions that are expected to arise depending on the sign of dispersion are to a large extent unknown.

This article is the first part of a study where we investigate in detail the bifurcation structure 
related to the different steady state solutions.
We focus in the present paper on the case of normal dispersion.
More specifically, we perform a stability analysis of the various solutions that can arise when a nonlinear WGM resonator is pumped in the normal dispersion regime. The control parameters of the stability map will be those that are the most easily accessible at the experimental level, that is, the pump power and the laser detuning relatively to the cavity resonance.

The plan of this article is the following.
In the next section, we will present the model that will be used to perform the stability analysis.
A short discussion will also be led to link the parameters of the model with the physical properties of the system under study. Section~\ref{equilibria} is devoted to the equilibria of the system and  their temporal stability. We then investigate the spatial bifurcations of the system in Sec.~\ref{spatialbifurcations}, and a global stability diagram is constructed in Sec.~\ref{bifurcationmap}. This stability study enables us to spot the basins of attraction of dark solitons, which are studied in Secs.~\ref{darkcavitysolitons} and \ref{coexistencesolitons}. We also analyze the effect of the magnitude of the dispersion parameter in Sec.~\ref{influencedispersion}.
The main findings of this study are synthetically resumed in the last section, which concludes this article.

\begin{figure}
\begin{center}
\includegraphics[width=8cm]{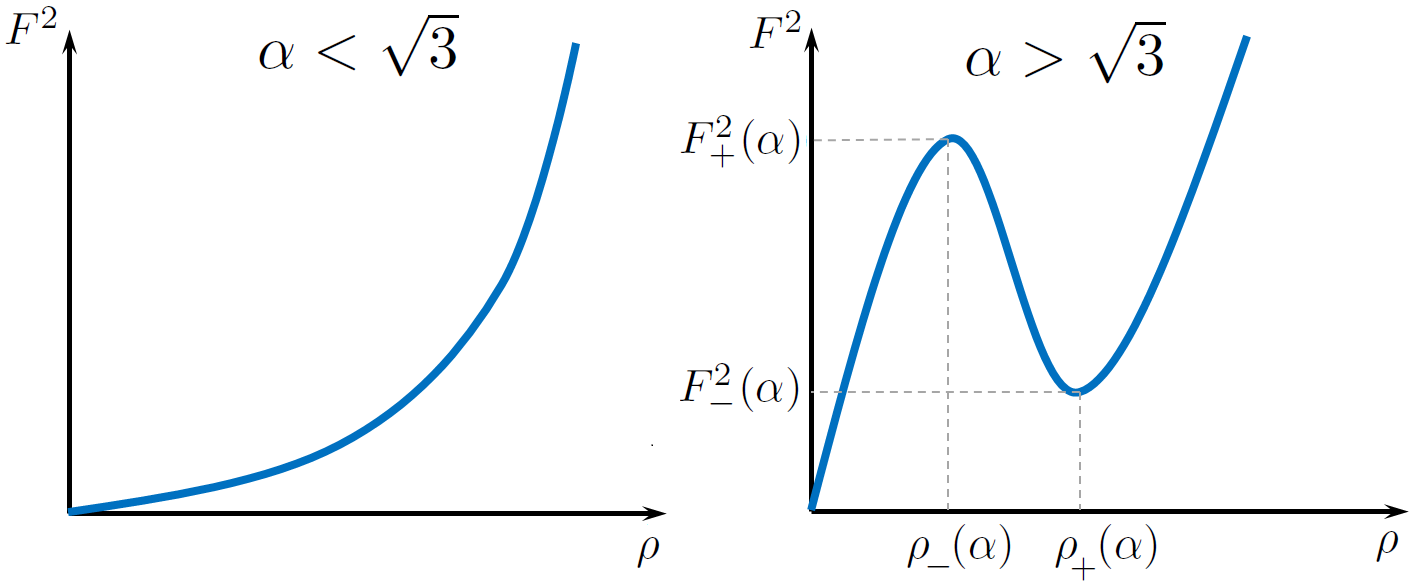}
\end{center}
\caption[Figure_equilibria]
{\label{Figure_equilibria} (Color online) 
Relationship between the number of solutions and the pumping strength $F$. There is always only one solution 
for $\alpha < \sqrt{3}$, but for $\alpha > \sqrt{3}$, there is a range if pumping strengths $F$ for which there are three solutions. }
\end{figure}

\section{The model}
\label{model}

The system under study is a WGM disk pumped by a CW pump laser radiation via evanescent coupling. 
Typical experimental set-ups are displayed in Fig.~\ref{Figure1}.
The understanding of the various phenomena of interest requires a sound knowledge of the eigenmode structure of WGM cavities.

Let us consider a disk of main radius $a$ and refraction index $n_0$ at the laser pump frequency $\Omega_0$. The simplest set of eigenmodes is the so-called fundamental family, which is characterized by a torus-like (or doughnut) spatial form inside the cavity. Each mode of this family can be unambiguously defined by a single eigennumber $\ell$, which can be interpreted as the number of internal reflections that a photon undergoes in that mode in order to perform a round-trip along the rim of the disk.

Let us now consider that the pumped mode is $\ell_0$. 
If we only consider the modes $\ell$ that are close to $\ell_0$, their eigenfrequencies can be
Taylor-expanded  as
\begin{eqnarray}
\omega_\ell = \omega_{\ell_0} + \zeta_1 (\ell - \ell_0) + \frac{1}{2}\zeta_2 (\ell - \ell_0)^2 
\label{freq_expansion}
\end{eqnarray}
where $\zeta_1 = c/ a n_0$ is the intermodal angular frequency (or free spectral range, FSR), with $c$ being the velocity of light in vacuum, while $\zeta_2$ stands for the second order dispersion which measures the unequidistance of the eigenfrequencies at the lowest order. 
We have here restricted ourselves to the second order in the Taylor expansion, but nothing forbids to 
consider higher-order dispersion terms if necessary. It is also interesting to note that the intracavity round-trip time is linked to the FSR by $T= 2 \pi / \zeta_1$.

\begin{table*}
\begin{center}
\begin{tabular}{|c|c|c|c|c|}
\hline 
\multicolumn{5}{|c|} {\textbf{}}\\
\multicolumn{5}{|c|}{\textbf{Eigenvalues and reversible spatial bifurcations in the system}} \\
\multicolumn{5}{|c|}{\textbf{}}\\
\hline Denomination & Eigenvalues $(\lambda_{1,2};\lambda_{3,4})$ & Pictogram &  Bifurcation   & Location on Fig.~\ref{Figurebifnotatscale}     \\
\hline
                     &                 &      &     &   \\
Type~1      & $(\pm a; \pm b)$ & \eigrenegnegpospos{-2}{1}{1}{-0.8}{3}{1.5} &     &   \\
                     &                 &       &     &  \\
\hline
                     &                 &       &     &  \\
Type~2     & $(0;0)$ & \ofour{-2}{1}{1}{-0.8}{3}{3.7}{3.6}  & ${0}^4$    & ${\mathsf{a}}$ \\
                     &                 &      &     &   \\
\hline
                     &                 &       &     &  \\
Type~3    & $(\pm ia; \pm ib)$ & \eigimnegnegpospos{-2}{1}{1}{-0.7}{3}{1.5}  &     &  \\
                     &                 &       &     &  \\
\hline
                     &                 &       &     &  \\
Type~4     & $(\pm a; 0)$ & \eigzerorenegpos{-2}{1}{1}{-0.7}{3}{1.5}{1.5} & ${0}^2$ & ${\mathsf{B}}_2$, ${\mathsf{b}}$, ${\mathsf{C}}_1$, ${\mathsf{c}}$, ${\mathsf{C}}_2$   \\
                     &                 &       &     &  \\
\hline
                     &                 &       &     &  \\
Type~5      & $(0; \pm ib)$ & \eigzeroimnegpos{-2}{1}{1}{-0.7}{3}{1.5}{1.5}  & ${0}^2 (i \omega)$  & ${\mathsf{B}}_1$   \\
                     &                 &       &     &  \\
\hline
                     &                 &      &     &   \\
Type~6     & $(\pm a; \pm ib)$ & \eigreim{-2}{1}{1}{-0.7}{3}{1.5}  &     &  \\
                     &                 &      &     &   \\
\hline
                     &                 &       &     &  \\
Type~7    & $(\pm ia; \pm ia)$ & \eigimnegpos{-2}{1}{1}{-0.7}{3}{1.5}  & $(i \omega)^2$ & ${\mathsf{A}}_3$  \\
                     &                 &      &     &   \\
\hline
                     &                 &       &     &  \\
Type~8   & $(\pm a; \pm a)$ & \eigrenegpos{-2}{1}{1}{-0.7}{3}{4}  &     &   \\
                     &                 &       &     &  \\
\hline
                     &                 &      &     &   \\
Type~9     & $(a \pm ib; c \pm id)$ & \eigreimout{-2}{1}{1}{-0.7}{3}{1.5}  &     &  \\
                     &                 &      &     &   \\
\hline
\end{tabular}
\end{center}
\caption{Nomenclature and pictograms for the various sets of eigenvalues.
A set of four eigenvalues is attached to each equilibrium, and some classified bifurcations are attached to certain configurations of eigenvalues.
A dot stands for one (simple) eigenvalue, the cross 
corresponds to for a set of two degenerated eigenvalues (double non semi-simple eigenvalue), and a circled cross stands for a set of four degenerated eigenvalues (quadruple eigenvalue with a $4 \times 4$ Jordan bloc).}
\label{NomenclatureBifurcations}
\end{table*}

The eigenmodes that are sufficiently close to $\ell_0$ are characterized by the same modal linewidth $\Delta \omega_{\rm tot}$. More precisely, we have 
\begin{eqnarray}
\Delta \omega_{\rm tot} = \Delta \omega_{\rm in} + \Delta \omega_{\rm ext}  \, ,
\label{def_linewidths_Q}
\end{eqnarray}
where $ \Delta \omega_{\rm in,ext,tot} =  \omega_{\ell_0} / Q_{\rm in,ext,tot}$ are respectively the intrinsic, extrinsic (or coupling) and total linewidths, while the quality factors $Q$ are defined analogously. 
Interestingly, the modal linewidth $\Delta \omega_{\rm tot}$ can be viewed as a measure of the total losses of the resonator, since it is linked to the average photon lifetime $\tau_{\rm ph}$ as 
\begin{eqnarray}
\tau_{\rm ph} = \frac{1}{\Delta \omega_{\rm tot}}  \, .
\label{def_tau_ph}
\end{eqnarray}
The normalized complex slowly varying envelopes ${\cal A}_{\ell}$ of the various eigenmodes can be obtained using the modal expansion model proposed in ref.~\cite{YanneNanPRA}. The amplitudes were normalized such that $|{\cal A}_{\ell}|^2$ was the number of photons in the mode $\ell$.
The overall intracavity field ${\cal A}$ can be determined as a sum of the modal fields ${\cal A}_{\ell}$, and in~\cite{PRA_Yanne-Curtis}, a spatiotemporal Lugiato-Lefever formalism has been constructed in order to describe the dynamics of this total intracavity field.
In its normalized form, this corresponding equation is the following partial differential equation 
\begin{eqnarray}
\frac{\partial \psi}{\partial \tau}  =  - (1 +i\alpha) \psi + i |\psi|^2 \psi - i \frac{\beta}{2} \frac{\partial^2 \psi}{ \partial \theta^2 } +F \, ,
\label{final_eq_dimensionless}
\end{eqnarray}
where  
\begin{eqnarray}
\psi (\theta ,\tau)  = \sqrt{\frac{2 g_0}{\Delta \omega_{\rm tot}}} \sum_{\ell} {\cal A}_\ell^* (\tau)  e^{\left[i (\ell - \ell_0) \theta + i\frac{1}{2}\beta (\ell - \ell_0)^2 \tau  \right]}
\label{def_psi}
\end{eqnarray}
is the complex envelope of the total intracavity
field, where $\theta \in [-\pi,\pi]$ is the azimuthal angle along the circumference, and $\tau=  t/2 \tau_{\rm ph}$ is the dimensionless time, with $\tau_{\rm ph}$ being the photon lifetime in the coupled cavity defined in Eq.~(\ref{def_tau_ph}). 
It is important to note that this equation has \emph{periodic} boundary conditions, and that $\psi$ represents the intracavity fields dynamics in the \emph{moving frame}.

The other dimensionless parameters of this normalized LLE are the frequency detuning
\begin{eqnarray}
\alpha = - \frac{2 (\Omega_0 -\omega_{\ell_0})}{\Delta \omega_{\rm tot}} \, ,
\label{def_alpha}
\end{eqnarray}
where $\Omega_0$ and $\omega_{\ell_0}$ are respectively the angular frequencies of the pumping laser and the cold-cavity resonance, and the overall dispersion parameter 
\begin{eqnarray}
\beta = - \frac{2 \zeta_2}{\Delta \omega_{\rm tot}} \, .
\label{def_beta}
\end{eqnarray}
Note that the anomalous GVD regime is defined by $\beta<0$ while normal GVD corresponds to $\beta>0$. 
Finally, using coupling formalism presented in ref.~\cite{Haus_coupling} and used in ref.~\cite{JSTQE} for optical resonators,
the dimensionless external pump field intensity can be explicitly defined as:
\begin{equation}
  F = \sqrt{\frac{8 g_0 \Delta \omega_{\rm ext}}{\Delta \omega_{\rm tot}^3} 
        \, \frac{P}{\hbar \Omega_0} } \, ,
  \label{eq:Power}
\end{equation}
where $P$ is the intensity (in W) of the laser pump at the input of the
resonator. The nonlinear gain $g_0$ is equal to $n_2 c \hbar \Omega_0^2/n_0^2 V_0$ where
$n_0$ and $n_2$ are respectively the linear and nonlinear refraction indices of the bulk material, and $V_0$ the effective volume of the pumped mode.
Note that since $F$ is real-valued and positive, the optical phase reference is arbitrarily set by this pump radiation for all practical purpose. \\

It is useful to give some orders of magnitude in relation to the model of Eq.~(\ref{final_eq_dimensionless}).
We consider the case of mm-size WGM resonators pumped around $1550$~nm.
The intermodal frequencies $\zeta_1/2 \pi$ are of the order of $10$~GHz, which correspond to round-trip times of $100$~ps and eigennumbers $\ell_0 \simeq \omega_{\ell_0} / \zeta_1 \sim 10^4$.
On the other hand, the dispersion parameter $\zeta_2$ is typically of the order of $1$~kHz or less in absolute value.
The $Q$ factors are typically of the order of $10^9$, so that the corresponding photon lifetimes $\tau_{\rm ph}$ are of the order of $1$~$\mu$s, and the modal linewidths are of the order of $100$~kHz. The pump power typically varies from $10$~mW to $1$~W.
These parameters can be easily translated to those of the LLE: the frequency detuning $\alpha$ can be linearly scanned at the experimental level to any value, but relevant values typically range from $-5$ to $5$ since out-resonance pumping occurs as soon as $|\alpha| >1$. The absolute value 
$|\beta|$ of the dispersion parameter is $10$ to $1000$~times smaller than unity. 
The pumping term $F$ will typically range between $0$ and $10$. Accordingly, the normalized intracavity field will also have an order of magnitude $|\psi|^2 \sim 1$.

We will use the spatiotemporal Lugiato-Lefever formalism represented by Eq.~(\ref{final_eq_dimensionless})
in order to investigate the various steady-state solutions that can emerge in the system. 
The numerical simulations will be performed using the split-step Fourier algorithm. 
It is noteworthy that this algorithm inherently assumes periodic boundary conditions because it is based on the fast Fourier transform
(FFT). It is therefore a very fast and efficient simulation method in our case where the periodic conditions are indeed periodic~\cite{PRA_Yanne-Curtis}.

\begin{figure*}
\begin{center}
\includegraphics[width=17cm]{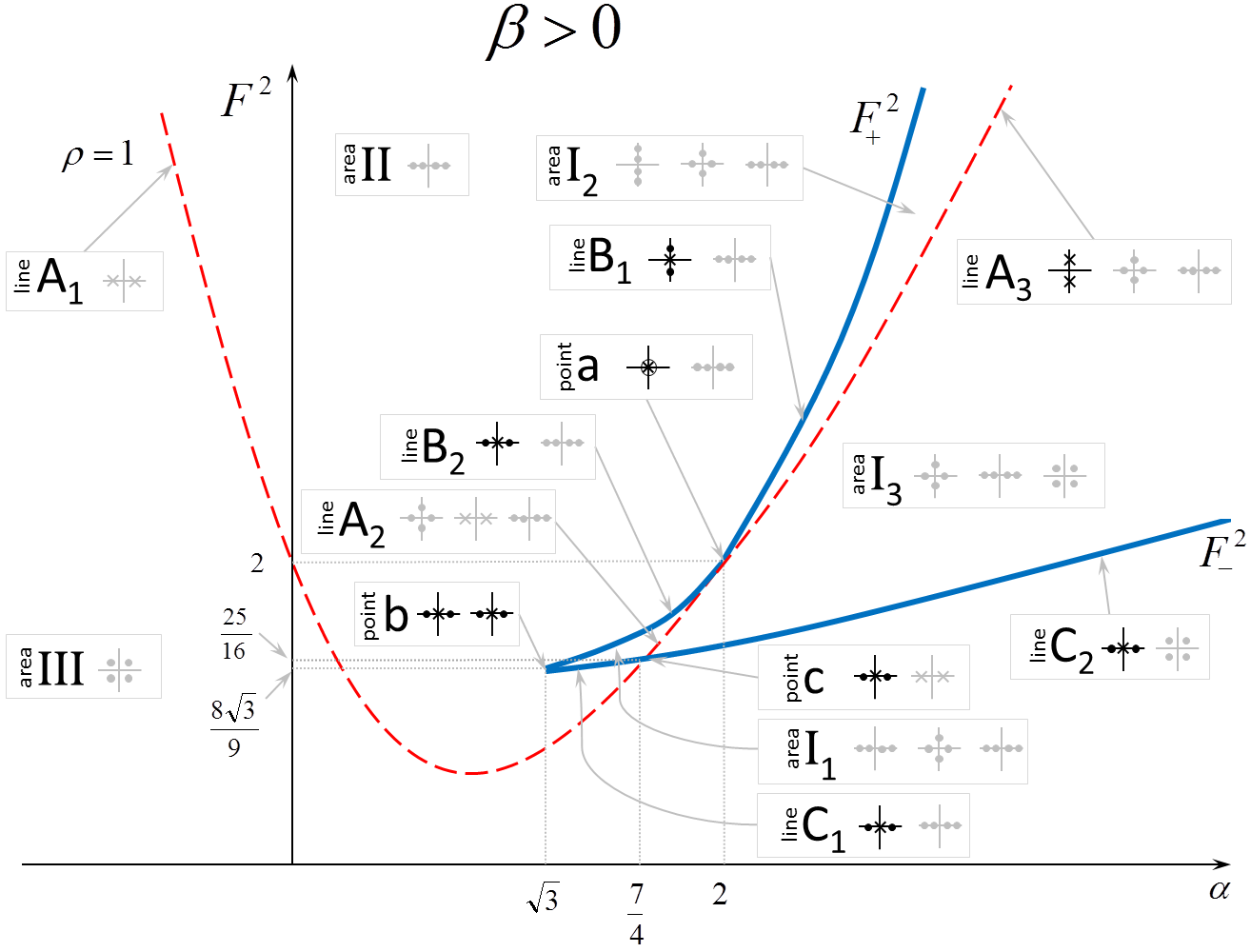}
\end{center}
\caption[Figurebif]
{\label{Figurebifnotatscale} (Color online) Bifurcation diagram (not at scale).
The areas are labelled using Roman numbers (I, II, and III), and area I is subdivided in three sub-areas (I$_1$, I$_2$, and I$_3$.
The lines are labelled using capital letters, with  line ${\mathsf{A}}$ standing for the limit $\rho = 1$ (dashed line in the figure), ${\mathsf{B}}$ standing for $F^2_+ (\alpha)$, and ${\mathsf{C}}$ standing for $F^2_- (\alpha)$. These lines can also be subdivided as ${\mathsf{A}}_1$, ${\mathsf{A}}_2$, etc.
The points are labelled using low-case letters (${\mathsf{a}}$ and ${\mathsf{b}}$). 
It is important to remember that the system has three equilibria in the area I (between the two thick lines $F^2_\pm$), and has only one equilibrium in areas II and III. Therefore, there is a set of three quadruplets of spatial eigenvalues in area I, two quadruplets in the boundaries  $F^2_\pm$, and one quadruplet outside. The eigenvalue pictograms are in black when they induce a bifurcation, and in grey when they do not. 
}
\end{figure*}

\section{Equilibria and their temporal stability}
\label{equilibria}

In this section, we aim to find the various equilibria of the system and determine their temporal stability.

All equilibria $\psi_{\rm e}$ are obtained from Eq.~(\ref{final_eq_dimensionless}) by setting all the derivatives to zero, thereby yielding 
\begin{eqnarray}
 F^2 = [1+ (\rho^2 - \alpha )^2] \rho^2   \equiv G (\alpha,\rho) \, ,
\label{equilibrium}
\end{eqnarray}
which is a cubic polynomial equation in $\rho = |\psi_{\rm e}|^2$.
It is well-known that this equation has one, two or three real-valued solutions depending on the parameters $\alpha$ and $F$.
Multiple solutions may arise in a polynomial equation when it has local extrema; in our case, this condition requires the existence of critical values of $\rho$ such that the partial derivative  
\begin{eqnarray}
\frac{{\partial G}}{\partial \rho} = 3 \rho^2 - 4 \alpha \rho + \alpha^2 + 1 
\label{dgdrho}
\end{eqnarray}
is null. This condition yields a quadratic equation with a discriminant equal to $4 (\alpha^2 - 3)$: therefore, if $|\alpha| < \sqrt{3}$, there are no such critical values for $\rho$ while for $|\alpha| \geq \sqrt{3}$,these critical values are 
\begin{eqnarray}
\rho_\pm (\alpha) = \frac{2 \alpha \pm \sqrt{\alpha^2 - 3}}{3}  \, . 
\label{rho_pm}
\end{eqnarray}
and the corresponding pumping terms are 
\begin{eqnarray}
F^2_\pm (\alpha) &=& G[\alpha, \rho_\pm (\alpha)]   \\ 
                 &=& \frac{2 \alpha \mp \sqrt{\alpha^2 - 3}}{3}
                  \left[1 + \Bigg(\frac{\sqrt{\alpha^2 - 3} \pm \alpha}{3} \Bigg)^2 \right] \, . \nonumber
\end{eqnarray}
Hence, when $\alpha> \sqrt{3}$, there exists a range of pumping power $F^2 \in ]F^2_-(\alpha), F^2_+(\alpha)[$ such that  there are three equilibria $\rho_1 < \rho_2 < \rho_3$. On the one hand, it can be shown that if these solutions are perturbed in the temporal domain, the extremal solutions $\rho_1$ and $\rho_3$ are always stable while the intermediate solution $\rho_2$ is always unstable. This is the well known theory of hysteresis in dynamical systems with cubic nonlinearity.  
On the other hand, outside the interval $[F^2_-(\alpha), F^2_+(\alpha)]$, there is a unique equilibrium which is always stable. The same conclusion applies as well for every value of $F^2$ whenever  $\alpha < \sqrt{3}$. 
The intermediate case were two solutions exist precisely corresponds to the boundary lines $F^2_-(\alpha)$ and 
$F^2_+(\alpha)$.
Figure~\ref{Figure_equilibria} shows how the multiple-equilibria states emerge as the value of the cavity detuning $\alpha$, 
increases. This analysis is actually equivalent to the one that was performed in ref.~\cite{YanneNanPRA} with the modal expansion model, when the time dynamics was disregarded (temporal derivative set to zero).



\begin{figure}
\begin{center}
\includegraphics[width=6cm]{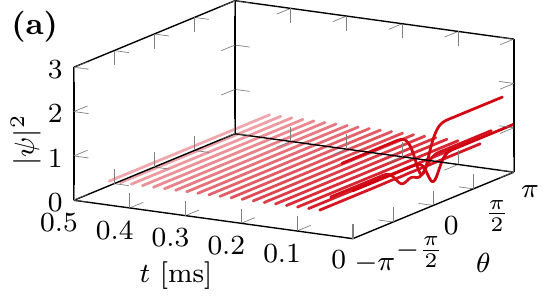}
\includegraphics[width=6cm]{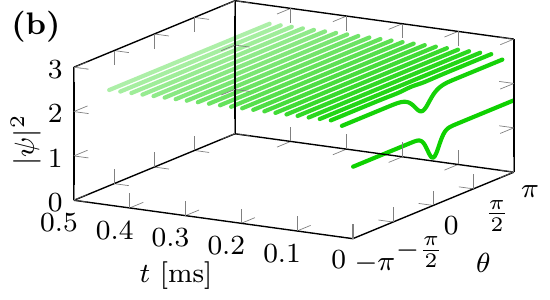}
\includegraphics[width=6cm]{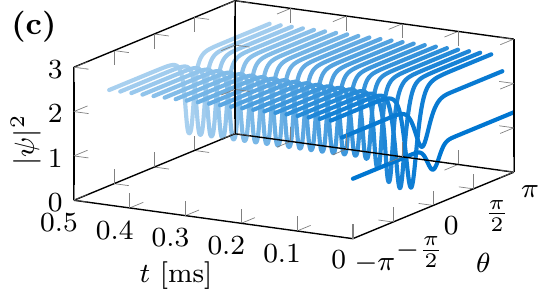}
\end{center}
\caption[Formation_soliton]
{\label{Formation_soliton} (Color online) Numerical simulation of the temporal dynamics of a pulse-like perturbation in the normal dispersion regime. The parameters are $\alpha = 2.5$, $\beta = 0.0125$ and $F^2 = 2.61$.
(a) Initial condition leading to a constant solution $\rho_1 = |\psi_{\rm down}|^2$.
(b) Initial condition leading to a constant solution $\rho_3 = |\psi_{\rm up}|^2$.
(c) Initial condition leading to the formation of a stable dark soliton.
}
\end{figure}

%

%

\section{Spatial bifurcations}
\label{spatialbifurcations}

The objective of a spatial bifurcation study is to investigate the various stationary solutions of the system as a function of the parameters.
The full study  requires at least the calculation of the relevant normal forms  around all the critical points and lines of the system. 
This task is indeed very complex, and can be circumvented by a simpler approach which can still provide insightful information about the spatial stability of the various solutions.

\begin{figure}
\begin{center}
\includegraphics[width=6cm]{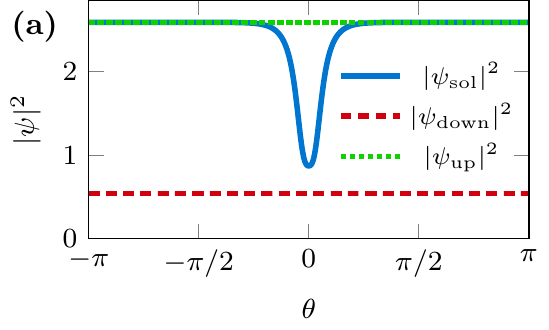}
\includegraphics[width=5cm]{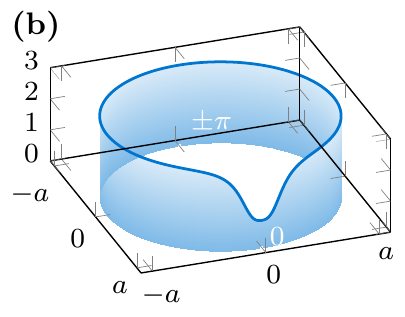}
\includegraphics[width=6cm]{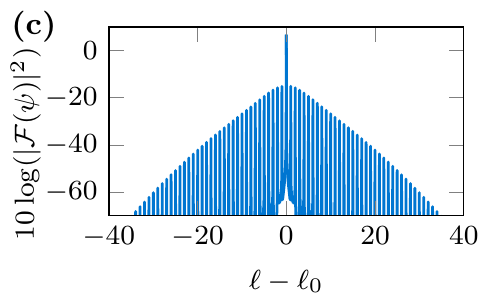}
\end{center}
\caption[Finalsoliton]
{\label{Finalsoliton} (Color online) 
(a) Asymptotic steady state profiles of the temporal dynamics presented in Fig.~\ref{Formation_soliton}.
Note that the power profile $|\psi_{\rm sol}|^2$ of the dark soliton lies between the ``up'' and
``down'' solutions.
(b) 3D representation of a dark soliton of Fig.~\ref{Formation_soliton}(c).
(b) Corresponding Kerr comb obtained using the Fast Fourier transform (FFT).
    The separation between the teeth corresponds to 1~FSR.
}
\end{figure}

We start with setting the temporal derivative to zero
and we rewrite the original Eq.~(\ref{final_eq_dimensionless}) as
\begin{eqnarray}
\frac{\partial^2 \psi_r}{\partial \theta^2} &=& \frac{2}{\beta} [ (\psi^2_r + \psi^2_i - \alpha) \psi_r - \psi_i] \\
\frac{\partial^2 \psi_i}{\partial \theta^2} &=& \frac{2}{\beta} [ (\psi^2_r + \psi^2_i - \alpha) \psi_i + \psi_r - F]
\label{eqs_stab_spatial}
\end{eqnarray} 
where $\psi = \psi_r + i \psi_i$, with $\psi_r$ and $\psi_i$ being respectively the real and complex parts of $\psi$. 
If we introduce the intermediate variable 
\begin{eqnarray}
\phi_{r,i} = \frac{\partial \psi_{r,i}}{\partial \theta} \, ,
\label{def_phi_ri}
\end{eqnarray}
then Eq.~(\ref{eqs_stab_spatial}) can be rewritten under the form of a four dimensional flow
\begin{eqnarray}
\frac{\partial \psi_r}{\partial \theta} &=& \phi_r \\
\frac{\partial \phi_r}{\partial \theta} &=& \frac{2}{\beta} ( \psi^3_r + \psi^2_i \psi_r - \alpha \psi_r - \psi_i)\\
\frac{\partial \psi_i}{\partial \theta} &=& \phi_i\\
\frac{\partial \phi_i}{\partial \theta} &=& \frac{2}{\beta} ( \psi^2_r \psi_i + \psi^3_i - \alpha \psi_i + \psi_r - F) \, .
\label{4D_flow_eqs_stab_spatial}
\end{eqnarray}

The matrix of the linearized system around an equilibrium $\psi_{\rm e} = \psi_{{\rm e},r} + i \psi_{{\rm e},i}$ is
 \begin{eqnarray}
\mathbf{J} =
            \begin{bmatrix}
                   0 & 1 & 0 & 0 \\
\frac{2}{\beta} ( 3 \psi_{{\rm e},r}^2\!+\! \psi_{{\rm e},i}^2 \!-\! \alpha ) &0& \frac{2}{\beta} (2 \psi_{
{\rm e},r} \psi_{{\rm e},i}\!-\! 1) & 0\\
0&0&0&1 \\
\frac{2}{\beta} (2 \psi_{{\rm e},r} \psi_{{\rm e},i}\!+\!1) &0&\frac{2}{\beta}(\psi_{{\rm e},r}^2\!\!+\! 3 \psi_{{\rm e},i}^2 \!-\! \alpha) & 0
            \end{bmatrix}
\end{eqnarray}
and the eigenvalues $\lambda$ of this Jacobian matrix obey the characteristic equation
\begin{eqnarray}
\lambda^4 \!- \!\frac{4}{|\beta|}(2 \rho\! - \! \alpha) \lambda^2 \!+ \!\frac{4}{|\beta|^2} (3 \rho^2 \! - \!4 \alpha \rho \!+\! \alpha^2 \!+ \!1) =0 \, .
\label{polynom}
\end{eqnarray} 
Since we consider here the case of normal dispersion, we have $\beta > 0$ but 
we keep a notation in $|\beta|$ in order to facilitate the comparison with the anomalous dispersion case where $\beta <0$.

Equation~(\ref{polynom}) is quadratic in $\lambda^2$: hence, we always have four eigenvalues which are either pairwise opposite (when real-valued) or pairwise conjugated (when complex-valued); it is also important to note that there is a quadruplet of eigenvalues \emph{for each solution}. We will therefore have $4$~eigenvalues in the area of the $\alpha$--$F^2$ plane where there is a single equilibrium,  $12$~eigenvalues in the hysteresis area where there are three equilibria, and $8$~eigenvalues in the boundary lines where there are two solutions.

The nature (complex, real, or pure-imaginary) of the eigenvalues is decided by the sign of the discriminant $\Delta = 16 (\rho^2 - 1)$, that is, by the comparative value of the equilibrium $\rho$ with regards to 1.
We hereafter analyze in detail the nature of the eigenvalues as a function of the sign of this discriminant, which is decided by $\rho>1$,  $\rho=1$, or  $\rho<1$.

\subsection{First case: $\rho>1$}

In this case the paired solutions obey 
\begin{eqnarray}
\lambda^2 \!=\!\frac{2}{|\beta|} [2 \rho \!-\! \alpha \! \pm \! \sqrt{\rho^2 \!-\!1}]   \, .
\label{sol_discr_pos}
\end{eqnarray}
The product of these paired solutions is equal to $3 \rho^2 \! -\! 4 \alpha \rho \! +\! \alpha^2 \!+ \!1 $, that is, to $\partial G /\partial \rho$ as defined in Eq.~(\ref{dgdrho}).
 
There are three sub-cases depending on the sign of this function:
\begin{itemize}
\item If $\partial G /\partial \rho >0$, 
 the eigenvalues can be written
  as   $(\lambda_{1,2};\lambda_{3,4})=(\pm a; \pm b)$ if $2 \rho - \alpha >0$  (this sub-case is referred to as Type~1); 
  as   $(\lambda_{1,2};\lambda_{3,4})=(0; 0)$ if $2 \rho - \alpha =0$ (Type~2), 
and as $(\lambda_{1,2};\lambda_{3,4})=(\pm ia; \pm ib)$ if $2 \rho - \alpha <0$ (Type~3).
\item If $\partial G /\partial \rho=0$, 
the eigenvalues can be written as 
      $(\lambda_{1,2};\lambda_{3,4})=(\pm a; 0)$ if $2 \rho - \alpha >0$ (Type~4); 
and as $(\lambda_{1,2};\lambda_{3,4})=(0; \pm ib)$ if $2 \rho - \alpha <0$ (Types~5). These sub-cases degenerates to a Type~2 eigenvalue when $2 \rho - \alpha =0$.
This sub-case mainly corresponds to solutions lying on the boundaries 
$F^2_\pm (\alpha)$.
\item If $\partial G /\partial \rho <0$, 
the eigenvalues have the form
      $(\lambda_{1,2};\lambda_{3,4})=(\pm a; \pm ib)$. This sub-case is referred to as an eigenvalue of Type~6.
\end{itemize}

\begin{figure*}
\begin{center}
\includegraphics[width=16cm]{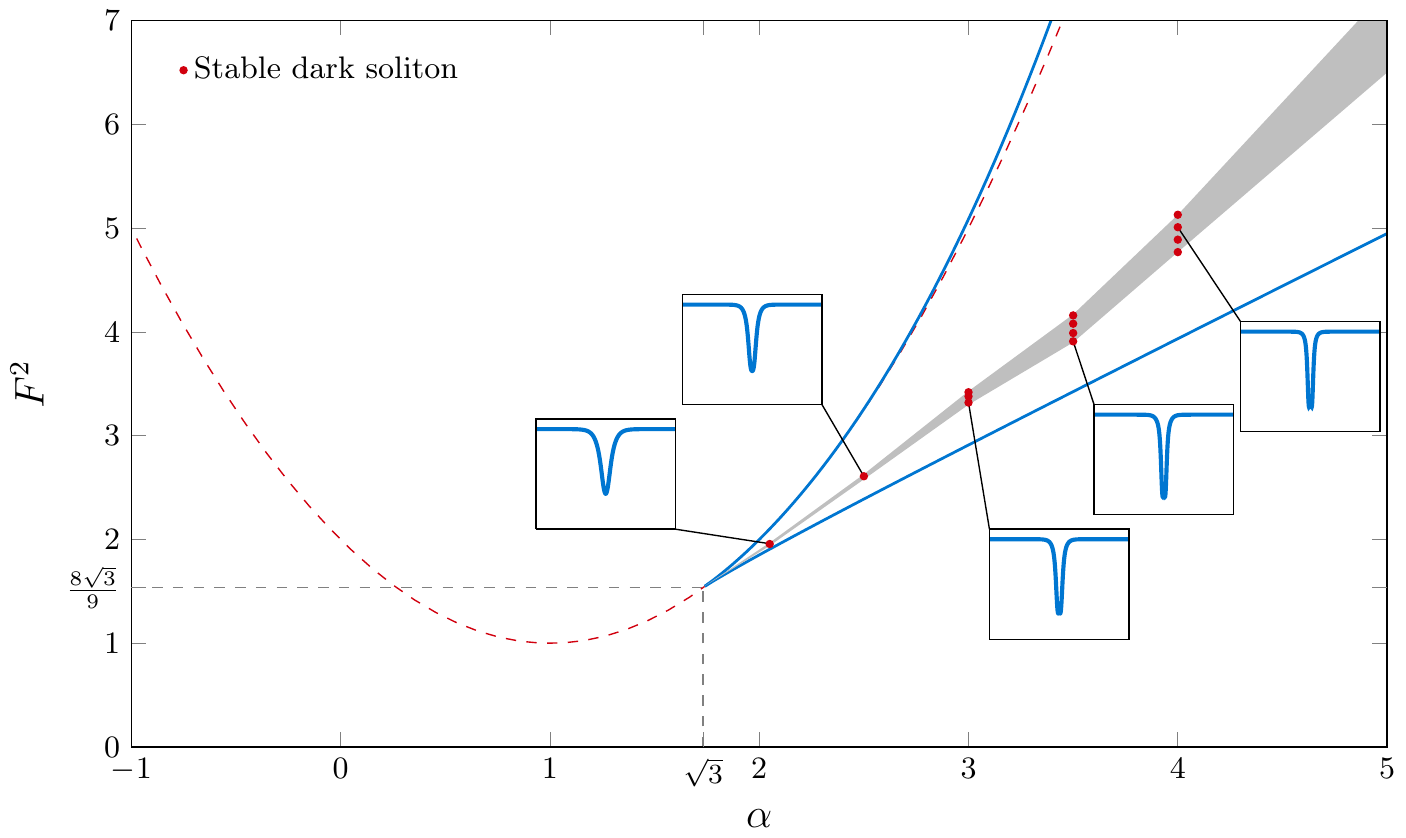}
\end{center}
\caption[Figurebifatscale]
{\label{Figurebifatscale} (Color online) Bifurcation diagram at scale, showing the parameters leading to various stationary solutions. Dark cavity solitons can be observed in a thin band laying within the three-equilibria area.
Out of this thin band (here approximated with straight lines), our numerical simulations have only evidenced convergence towards stable equilibria (flat solutions).}
\end{figure*}

\subsection{Second case:  $\rho=1$}

Here, the characteristic equation has a double-root
\begin{eqnarray}
\lambda^2 \!=\!\frac{2}{|\beta|} [2  \!-\! \alpha ]   \, .
\label{sol_discr_null}
\end{eqnarray}
Hence, the eigenvalues can be written 
as $(\lambda_{1,2};\lambda_{3,4})=(\pm ia; \pm ia)$ when $\alpha >2$ (Type~7);
as $(\lambda_{1,2};\lambda_{3,4})=(0;0)$ when $\alpha =2$ (Type~2); 
and as $(\lambda_{1,2};\lambda_{3,4})=(\pm a; \pm a)$ when $\alpha <2$ (Type~8).

\subsection{Third case:  $\rho<1$}

This case corresponds to the situation where the eigenvalues are complex:
\begin{eqnarray}
\lambda^2 \!=\!\frac{2}{|\beta|} [2 \rho \!-\! \alpha \! \pm \! i\sqrt{1 \!-\!\rho^2}]   \, .
\label{sol_discr_neg}
\end{eqnarray}
This kind of eigenvalues will be referred to as of Type~9, and they have the explicit form
$(\lambda_{1,2};\lambda_{3,4})=( a \pm ib ; c \pm id)$.

\section{Bifurcation map}
\label{bifurcationmap}

The stability analysis developed in
Sec.~\ref{spatialbifurcations} enables us to obtain a bifurcation map, which is presented in Fig.~\ref{Figurebifnotatscale}.

It is important to note that in our case, the system has a $\theta \rightarrow -\theta$ symmetry: as a consequence, the spatial bifurcations that are arising in the system are necessarily \textit{reversible}.
Such reversible bifurcations have been studied in detail in the fourth Chapter of ref.~\cite{Mariana_Iooss}, where a normal form characterization is provided as well. This reversibility is essential because it facilitates the study of the bifurcations. Another consequence of this symmetry is that the eigenvalue spectrum is symmetrical relatively to the imaginary axis, which in our system comes along with a symetry relatively to the real axis. This is why these eigenvalues are structurally similar to those of an Hamiltonian system. 

We hereafter list the four reversible bifurcations that can be identified in Fig.~\ref{Figurebifnotatscale}.

\begin{itemize}

\item {${0}^2$ bifurcation}: 
The ${0}^2$ bifurcation (also referred to as Takens-Bogdanov bifurcation)  arises when a quadruplet of eigenvalues
is of Type~4.
This situation is witnessed in our system along  the lines ${\mathsf{B}}_2$, ${\mathsf{C}}_1$, and ${\mathsf{C}}_2$.
Depending on the system under study, both periodic and localized stationary solutions can eventually be sustained at the vicinity of this bifurcation.

\item {${0}^2 (i \omega)$ bifurcation}:
This bifurcation corresponds to a quadruplet of Type~5, and it is present in our system as the line ${\mathsf{B}}_1$. 
Along this bifurcation, possible stationary states are periodic and quasiperiodic solutions.
However, localized solutions are typically unstable near this bifurcation.

\item {$(i \omega)^2$ bifurcation}:
This bifurcation is sometimes referred to as the ``$1:1$ resonance'' or the ``Hamiltonian Hopf'' bifurcation.
It arises when a quadruplet of eigenvalues is of Type~7, and in Fig.~\ref{Figurebifnotatscale}, it corresponds to the line ${\mathsf{A}}_3$. 
Typical solutions around this bifurcation eventually include quasiperiodic, periodic and localized solutions.

\item {${0}^4$ bifurcation}:
This is a co-dimension~2 bifurcation, arising when a quadruplet of eigenvalues degenerates to the origin (Type~2).
This situation only arises on the point ${\mathsf{a}} \equiv (2,2)$, around which a very large variety of dynamics can be observed a priori.

\end{itemize}

Since some eigenvalues are located on the imaginary axis in the pictograms of Type~3 and~6, they also indeed correspond to bifurcations, respectively referred to as $(i \omega_1)(i \omega_2)$ and to $(i \omega)$. 
However, the reversibility of our system forces these eigenvalues to stay on the imaginary axis, so that these bifurcations are not dynamically relevant in our system (non-respect of the transversality condition). This is why they are not highlighted in 
Table~\ref{NomenclatureBifurcations}.

It is important to note that in this regime of normal dispersion, the boundary line $\rho = |\psi|^2 =1$ \textit{does not} correspond to a bifurcation when $\alpha <2$. In particular, for $\alpha <\sqrt{3}$, we have shown earlier that there is only one equilibrium, 
which is necessary stable. Hence, the bifurcation analysis indicates that increasing the pump power in that case does not lead 
to any modification since areas I and III do not differ structurally: this is why Kerr comb generation is impossible for $\alpha < \sqrt{3}$ in the regime of normal dispersion.
Actually, since most of the time only in-resonance pumping is investigated ($|\alpha| < 1$), it has been thought
for long time that Kerr comb generation was absolutely impossible in the regime of normal dispersion.
Normal GVD Kerr combs have in fact been numerically evidenced only very recently using a modal expansion model~\cite{Matsko_Normal}.

However, this bifurcation analysis indicates that non-trivial solutions might exist around the bifurcation lines we
have identified. We will show in the next sections that dark solitons can arise in the system and can lead to complex patterns in both the time and spectral domains.

\begin{figure}
\begin{center}
\includegraphics[width=6cm]{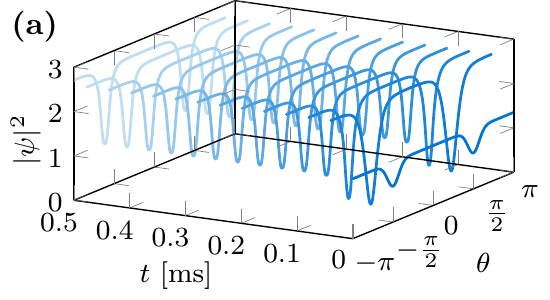}
\includegraphics[width=5cm]{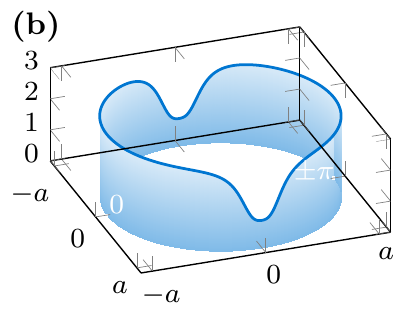}
\includegraphics[width=6cm]{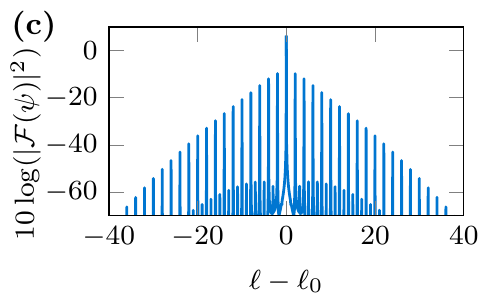}
\end{center}
\caption[twosolitons]
{\label{twosolitons} (Color online) 
(a) Temporal dynamics leading to a the formation of two solitons in the cavity. The pulses are evenly spaced in this case (separated by $\pi$ and dips at $\pm \pi/2$). The parameters are those of Fig.~\ref{Formation_soliton}, which are $\alpha = 2.5$, $\beta = 0.0125$ and $F^2 = 2.61$, so that each pulse is identical to the isolated pulse presented in Figs.~\ref{Formation_soliton}(c) and~\ref{Finalsoliton}(b) at $\tau=+\infty$.
(b) 3D representation. 
(c) Corresponding Kerr comb. Note that the separation between the teeth corresponds to $2$~FSRs. 
The single-FSR ``sub-comb'' around $-60$~dB arises from the fact that the two pulses are not perfectly identical and perfectly equidistant.
}
\end{figure}

\section{Dark cavity solitons}
\label{darkcavitysolitons}

Dark solitons are stable localized structures characterized by a hole in a finite background
(see review article~\cite{DarSolitonReview}).
In our system, the formation of these solitons is explicited in Fig.~\ref{Formation_soliton} where the 
temporal dynamics of the intracavity field $\psi$
is displayed. 
It can be seen that depending on the initial conditions, the final steady state of the
field can either be a stable equilibrium (that is, a stable solution of Eq.~(\ref{equilibrium})) or a dark soliton.
We had explained earlier that in the three-solutions area, the intermediate solution is
generally unstable while the extremal solutions are stable.
We can see in Fig.~\ref{Formation_soliton}(a) and~(b) that for the same sets of parameters,
the systems may converge to the downmost or
uppermost steady-state solution depending on the initial conditions.
However, we can show
that a dark soliton can appear as a stable and robust
solution which is intermediate between the asymptotic levels of the two extremal
steady-states, as evidenced in Fig.~\ref{Finalsoliton}(a).
The typical spectrum of a dark soliton is displayed in Fig.~\ref{Finalsoliton}(c), and shows a typical triangular-like decrease of modal power away from the pump.
It is noteworthy that because there is only one pulse inside the cavity, the corresponding comb has a single-FSR spacing.
These solitons have already been observed experimentally~\cite{IEEE_PJ}.

\begin{figure}
\begin{center}
\includegraphics[width=6cm]{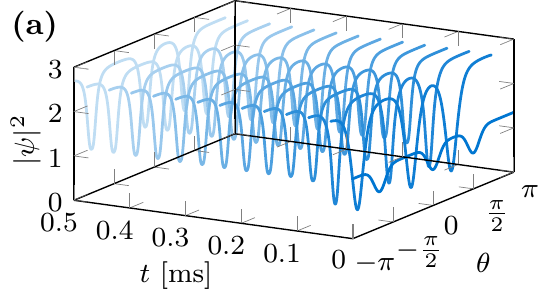}
\includegraphics[width=5cm]{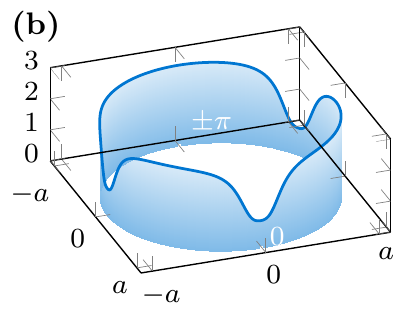}
\includegraphics[width=6cm]{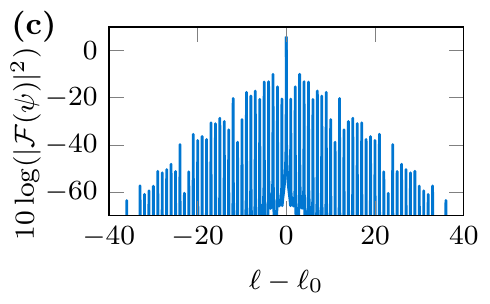}
\end{center}
\caption[threesolitons]
{\label{threesolitons} (Color online) 
Coexistence of three solitons in the cavity with $\alpha = 2.5$, $\beta = 0.0125$ and $F^2 = 2.61$ 
(same as in Fig.~\ref{Formation_soliton}). 
(a) Temporal dynamics leading to a the formation of three solitons in the cavity.
Note that the solitons are not evenly spaced (dips at $-2 \pi/3$, $0$, and  $\pi/2$).
(b) 3D representation. 
(c) Corresponding Kerr comb. The separation between the teeth is equal to single-FSR, but because of the unequidistance of the pulses (which are however identical in shape), the comb looks ``irregular". A three-FSR comb would have been obtained if the pulses (which are identical) were also equidistant and separated by $2 \pi /3$.  
}
\end{figure}

Since dark solitons are intermediate solutions between the stable equilibria in our system, they do exclusively appear in the area I, where these three equilibria actually exist. However, the existence of multiple equilibria is a necessary but not sufficient condition for the emergence of stable dark solitons. As displayed in Fig.~\ref{Figurebifatscale}, our numerical simulations show that they can be observed only for a restricted range of parameters laying within a thin band inside the three-equilibria area. 
A direct consequence of this observation is that no Kerr comb generation is a priori possible in the areas~I and~II characterized by single equilibria. 
Because fairly large (out of resonance) detunings are required for the emergence of dark solitons in Fig.~\ref{Figurebifatscale}, Kerr comb generation is very difficult to obtain experimentally in the normal GVD regime.
These combs are also likely to be only weakly stable whenever observed.  
Numerical simulations have also indicated that the solitons seem to loose their stability as $\alpha$ becomes larger.
Further studies are required in order to elucidate their dynamical properties in this asymptotic regime (note however that $\alpha > 5$ is generally not easy to achieve under normal circumstances, since the out-of-resonance pumping starts at $\alpha =1$).

It is also important to note that the emergence of dark solitons depends on the initial conditions. 
Actually, from an experimental point of view, dark solitons will not naturally arise 
from noise above a certain threshold. The most likely outcome in that case would be a convergence towards the nearest flat (constant) solution, which is $\rho_1 = |\psi_{\rm down}|^2$. 
Only a compact (but continuous) set of initial conditions $\psi (\theta, \tau=0)$ can lead to the dark soliton.
From this standpoint, Fig.~\ref{Finalsoliton} can be viewed as the result of multi-stability, with each stable solution
$\psi_{\rm down}$,  $\psi_{\rm up}$ and $\psi_{\rm sol}$ having its own basin of attraction.

\section{Coexistence of dark cavity solitons}
\label{coexistencesolitons}

The dark solitons that were investigated in the last section are localized dissipative structures: they do not see the boundary conditions and actually, they could be observed as well if the background support was extended to infinity.
Hence, multiple dark cavity solitons can co-exist in the cavity as long as they are far away from each other and can not ``feel'' each other's influence (they are not bounded). 

This situation for example presented in Figs.~\ref{twosolitons} where two solitons are excited inside the cavity with a double-pulse initial condition. In this case, the solitons have been evenly spaced (separation of $\pi$) and as a consequence, the corresponding Kerr comb has a two~FSR separation. 
A more complex case is presented in Fig.~\ref{threesolitons}, where three unevenly-spaced solitons are excited inside the cavity. The corresponding spectra has single-FSR separation but displays a very complex pattern. 
Indeed, if the pulses were equidistant, the Kerr comb would have displayed a three-FSR separation.
These multiple-FSR combs will generally appear whenever we have $N$ identical dark solitons separated by a $2 \pi /N$ angle.
However, this symmetry is fragile and will be broken by non-identical and/or non-equidistant pulses, yielding  non-trivial Kerr comb spectra. Of course, these spectra should not be referred to as ``chaotic'' or ``noisy'', since the corresponding time-domain solutions are perfectly periodic and deterministic.
 
It is worth noting that in the multiple-pulse regime, each pulse is identical to the solitary pulse presented in
Fig.~\ref{Finalsoliton}(b). Reaching the single- or multiple-soliton states only depends on the  initial condition. 
Therefore, these non-bounded multiple-soliton states emerge in the shaded area of the parameter space area in Fig.~\ref{Figurebifatscale}, exactly as single-pulse localized structures.

\begin{figure}
\begin{center}
\includegraphics[width=8cm]{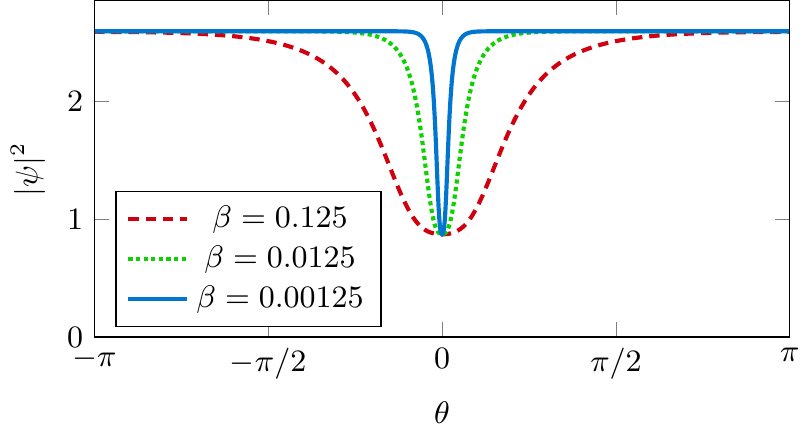}
\end{center}
\caption[threesolitons]
{\label{influenceofbeta} (Color online)
Evidence of the effect of $\beta$, showing how the soliton pulses become narrower as $\beta \rightarrow 0$. 
The detuning and pump  parameters $\alpha = 2.5$ and $F^2 = 2.61$ are fixed, while the dispersion parameter is logarithmically  varied as 
$\beta = 0.00125$  $\beta = 0.0125$ (same as in Fig.~\ref{Formation_soliton}), and $\beta = 0.125$.
}
\end{figure}

\section{Influence of the dispersion parameter $\beta$}
\label{influencedispersion}

We have analyzed so far the bifurcation map with regards to the two control parameters $\alpha$ and $F$.
The dispersion parameter $\beta$ does actually not have a direct influence on this map because we have focused on localized structures and we have disregarded the boundaries.
However, $\beta$ has a direct influence on the temporal profile of the pulses, and particularly on their widths.
Figure~\ref{influenceofbeta} shows that as the dispersion decreases, the pulse-width decreases as well.
It should be noted that from a physical  standpoint, higher-order dispersion has to be included into the Lugiato-Lefever model when second-order dispersion vanishes~\cite{PRA_Yanne-Curtis}. 

Since the pulses become narrower as $\beta \rightarrow 0$, it also becomes easier to excite a large number of solitons in the cavity, following the scenario explained in Sec.~\ref{coexistencesolitons}, and the corresponding Kerr comb spectra will also display increasingly complicated patterns.\\

\section{Conclusion}
\label{Conclusion}

In this article, we have performed a bifurcation analysis of Kerr comb generation in the normal dispersion regime based on the Lugiato-Lefever model.
The non-trivial solutions we have spotted are dark cavity solitons which can be stable in the three-equilibria area only.
In this regard, the single-equilibrium area (which encompasses the in-resonance pumping regime) appeared to be unsuitable for Kerr comb generation.
This analysis has shown that normal GVD Kerr combs can not arise straightforwardly from noise, but are excited for only certain initial conditions. We have also shown that the underlying dark solitons can coexist in the cavity as long as they do not interact.
It is noteworthy that we have only focused here on localized solutions, and eventual solutions that might depend on the finite-background boundary conditions have not been considered here.
Further studies are therefore necessary in order to explore exhaustively this parameter space with this additional constraint.

\section*{Acknowledgements}

The authors acknowledge financial support from the European Research Council
through the project NextPhase (ERC StG 278616).
They would also like to thank Mariana Haragus for very insightful comments and suggestions.
Authors would also like to acknowledge for the support
of the \textit{M\'esocentre de Calcul de Franche-Comt\'e}.


\begin{thebibliography}{99}
\bibitem{Review_Kerr_combs_Science} T. J. Kippenberg, R. Holzwarth, and S. A. Diddams,
 {\it Science} {\bf 322}, 555 (2011).
\bibitem{Nature_Ferdous} F. Ferdous,H. Miao, D. E. Leaird,	K. Srinivasan, J. Wang,
                          L. Chen,	L. T. Varghese	and A. M. Weiner,
                          \textit{Nature Photonics}~\textbf{5}, 770 (2011). 
\bibitem{Nature_Selectable_freq}  A. A. Savchenkov,	A. B. Matsko, W. Liang,	V. S. Ilchenko,	
                                    D. Seidel	and  L. Maleki, 
                                    \textit{Nature Photonics}~\textbf{5}, 293 (2011).
\bibitem{PRA_Scott2} S.~B. Papp and S.~A. Diddams, \emph{Phys. Rev. A} \textbf{84},
                     053833 (2011).
\bibitem{PRL_NIST} P.~Del'Haye, S.~B. Papp, and S.~A. Diddams, 
                    \emph{Phys. Rev. Lett.} \textbf{109}, 263901 (2012). 
\bibitem{Nature_Universal}     T. Herr,	K. Hartinger, J. Riemensberger,	C. Y. Wang,	
                               E. Gavartin, R. Holzwarth, M. L. Gorodetsky, and T. J. Kippenberg,
                               \textit{Nature Photonics}~\textbf{6}, 480 (2012).
\bibitem{Nature_Mid_IR} C. Y. Wang, T. Herr, P. Del'Haye, A. Schliesser, J. Hofer,	R. Holzwarth,	
                        T. W. Haensch, N. Picqu\'e, and T. J. Kippenberg, 
                        \emph{Nature Communications}~\textbf{4},  134~(2013).
\bibitem{Jove} A. Coillet, R. Henriet, K. P. Huy, M. Jacquot, L. Furfaro, I. Balakireva, L. Larger, and Y. K. Chembo,
                  to be published in~\emph{J. Vis. Exp.}, (2013).
\bibitem{Maleki_PRL_LowThres}
A.~A. Savchenkov, A.~B. Matsko, D.~Strekalov, M.~Mohageg, V.~S. Ilchenko, and L.~Maleki,
  \emph{Phys. Rev. Lett.}~\textbf{93}, 243905 (2004).
\bibitem{YanneNanPRL} Y. K. Chembo, D. V. Strekalov, and N. Yu,  {\it Phys. Rev. Lett.} {\bf 104}, 103902
 (2010).
\bibitem{YanneNanPRA} Y. K. Chembo and N. Yu,  {\it Phys. Rev. A} {\bf 82},
 033801 (2010).
\bibitem{GaetaOE}
I.~H. Agha, Y.~Okawachi, and A.~L. Gaeta, \emph{Opt. Express}~\textbf{17}, 16209 (2009). 
\bibitem{Matsko_OL_2}
A.~B. Matsko, A.~A. Savchenkov, W.~Liang, V.~S. Ilchenko, D.~Seidel, and
  L.~Maleki,  \emph{Opt. Lett.}~\textbf{36}, 2845 (2011).
\bibitem{PRA_Yanne-Curtis} Y.~K. Chembo and C.~R. Menyuk, 
                           \emph{Phys. Rev. A}~\textbf{87}, 053852 (2013).
\bibitem{Coen} S.~Coen, H.~G. Randle, T.~Sylvestre, and M.~Erkintalo, 
              \emph{Opt. Lett.}~\textbf{38}, 37 (2013). 
\bibitem{LL} L.~A. Lugiato and R.~Lefever, \emph{Phys. Rev. Lett.}~\textbf{58}, 2209 (1987).
\bibitem{IEEE_PJ} A. Coillet, I. Balakireva, R. Henriet, K. Saleh,
                  L. Larger, J. M. Dudley, C. R. Menyuk, and Y. K. Chembo,
                  to be published in~\emph{IEEE Photonics Journal}, (2013).
\bibitem{Haus_coupling} H. A. Haus, ``Waves and fields in optoelectronics'', Chapter~7,
         Prentice-Hall (1984).
\bibitem{JSTQE}   A. Coillet, R. Henriet, P. Salzenstein, K. Phan-Huy, L. Larger, and Y. K. Chembo,
                   \emph{IEEE Sel. Top. Quantum Electron.}~\textbf{19}, 6000112 (2013).      
\bibitem{Mariana_Iooss} M. Haragus and G. Iooss, ``Local Bifurcations, Center Manifolds, and Normal Forms in Infinite-Dimensional Dynamical Systems'', Springer (2010). 
\bibitem{Matsko_Normal} A. B. Matsko, A. A. Savchenkov, and L. Maleki,
                          \textit{Opt. Lett.}~\textbf{37}, 43 (2012).
\bibitem{DarSolitonReview} B. Luther-Davies and Y. S. Kivshar,
                          \textit{Physics Reports}~\textbf{298}, 81 (1998).
\end{thebibliography}
\end{document}